# Self-amplified Emission of Cherenkov Radiation from a Relativistic Particle Inside Layered Dielectric-filled Waveguide

<u>L.Sh. Grigoryan</u>, H.F. Khachatryan, S.R. Arzumanyan

Institute of Applied Problems in Physics, Yerevan 0014, Armenia

The radiation from a charged particle uniformly moving along the axis of cylindrical waveguide filled with semi-infinite layered dielectric that weakly absorbs the radiation is investigated. Expressions have been derived for calculation of the spectral distribution of total energy of radiation passing through the transverse section of waveguide far from the layered material during the whole time of particle travel.

The results of numerical calculations for emission of Cherenkov radiation (CR) in the layered material consisting of dielectric plates alternated with vacuum gaps are given. It was shown that in some special cases CR generated in each dielectric plate is found in the neighbouring plate together with the particle that generates there the new CR. As a result, in the zone of CR formation there constantly occurs the superposition of CR pulses emitted by the particle earlier. The interference of this kind may be followed either by the suppression, or the amplification of CR. The values of the parameters of problem under which the amplification (self-amplification) of CR at a separate waveguide mode have been determined. This radiation may prove to be many times as strong as CR in the waveguide filled with semi-infinite solid dielectric without vacuum gaps.

It is proposed to use this effect for amplification of the power of coherent CR from pico- and subpicosecond electron bunches [1,2]. In [3,4] a similar effect for transition radiation was experimentally studied. The difference is that in [3,4] a train of electron bunches was used instead of one charged particle, one plate was used instead of a stack of plates, and a special system of mirrors was used instead of the waveguide.